\documentclass[12pt]{article}
\usepackage{amsmath,amssymb,amscd,cite}
\newtheorem{thm}{Theorem}[section]
\newtheorem{prop}[thm]{Proposition}
\newtheorem{lem}[thm]{Lemma}

\newtheorem{defi}[thm]{Definition}
\newcommand{\pf}{{\bf Proof. \ }}
\newcommand{\qed}{\hfill $\Box$ \\}

\date{}
\begin{document}
\title{On Some Classes of $\mathbb{Z}_{2}\mathbb{Z}_{4}-$Linear Codes and their Covering Radius}
\author{K. Chatouh, K. Guenda, T.A.Gulliver and L. Noui}
\maketitle
\begin{abstract}
In this paper we define $\mathbb{Z}_{2}\mathbb{Z}_{4}-$Simplex and MacDonald
Codes of type $\alpha $ and $\beta $ and we give the covering radius of these
codes.
\end{abstract}
 \section{Introduction }
 Much research has concerned the construction of
the additive codes especially, Delsarte in 1973, who put the first definition of this codes \cite{Delsarte}. For more information concerning the $\mathbb{Z}_{2}\mathbb{Z}_{4}$-additive code the reader is invited to consult (see \cite{Bilal}, \cite{Fernandez}, \cite{Borges}, \cite{Pujol} and \cite{Kim}).
The simplex and MacDonald codes  over a finite fields and a finite rings are studied in several version of those articles. (see \cite{AL-Ashker}, \cite{Bhandari}, \cite{Colbourn} and \cite{Manish}). We will use these codes to define a new family of simplex and MacDonald codes is a concatenation of a binary and quaternary simplex and MacDonald codes.

The subject of this paper is the construction of simplex and MacDonald codes over $\mathbb{Z}_{2}\mathbb{Z}_{4}$ of type $\alpha$ and $\beta$ is also their covering radius. The paper is organized as follows. In Section 2, we recall some properties related to  $\mathbb{Z}_{2} \mathbb{Z}
_{4}$-additive codes. In Section 3, we calculated the covering radius of  $\mathbb{Z}_{2} \mathbb{Z}_{4}$-  repetition codes. In Section 4 and 5, we describe the construction of $\mathbb{Z}_{2} \mathbb{Z}_{4}$- simplex, MacDonald codes and we estimated theirs covering radius. In Section 6, We concluded a relation between the first order of Reed Muller codes and simplex codes over  $\mathbb{Z}_{2}\mathbb{Z}_{4}$. Finally in section 7, we give the binary version of these codes.

 \section{Preliminaries }
In this section based on some articles (see \cite{Bilal}, \cite{Fernandez} and \cite{Pujol}) to select General preliminaries which serve this search.

 Denote by $\mathbb{Z}_{2}$ and $\mathbb{Z}_{4}$ the rings of integers modulo
$2$ and modulo $4$, respectively. Let $\mathbb{Z}_{2}^{n}$ and $\mathbb{Z}%
_{4}^{n}$ denote the space of $n-$tuples over these rings. We say that a
binary code is any nonempty subset $C$ of $\mathbb{Z}_{2}^{n}$, and if that
subcode is a vector space then we say that it is a linear code. Similarly,
any nonempty subset $C$ of $\mathbb{Z}_{4}^{n}$ is called a linear
quaternary code.

 The codes considered here are subgroups of the space $\mathbb{Z}_{2}^{\gamma }\times \mathbb{Z}
_{4}^{\delta }$. A code C is $\mathbb{Z}_{2}\mathbb{Z}
_{4}$-additive if the set of coordinates can be partitioned into two subsets
$X$ and $Y$ such that the punctured code of $C$ by deleting the coordinates outside $X$ (respectively,
$Y$ ) is a binary linear code (respectively, a quaternary linear code). Let $C_{b}$ be the subcode of $C$ which contains all order two codewords and
let $κ$ be the dimension of $\left(C_{b} \right)_{X}$, which is a binary linear code. For the case $\gamma=0$, we will
write $k = 0$. Considering all these parameters, we will say that $C$ (or equivalently $C =\Phi(C)$
is of type $\left(\gamma,\delta;\lambda,\mu ;k\right) $.
The structure of these codes is given as follows. We write $v=\left( v_{1},v_{2}\right)\in \mathbb{Z}_{2}^{\gamma }\times \mathbb{Z}
_{4}^{\delta } $ where $v_{1}=\left(x_{1},\cdots,x_{\gamma} \right)\in \mathbb{Z}_{2}^{\gamma } $ and $v_{2}=\left(y_{1},\cdots,y_{\delta } \right)\in \mathbb{Z}_{4}^{\delta }$.

The code $C$ is a subgroup of $\mathbb{Z}_{2}^{\gamma }\times \mathbb{Z}
_{4}^{\delta }$, as such it is isomorphic to $\mathbb{Z}_{2}^{\lambda
}\times \mathbb{Z}_{4}^{\mu }$ for some $\lambda $ and $\mu $. We say that $
C $ is of type $2^{\lambda }4^{\mu }$ as a group. It follows that it has $
\left\vert C\right\vert =2^{\lambda +2\mu }$ codewords and the number of
order two codewords in $C$ is $2^{\lambda +\mu }$. We will take an extension
of the usual Gray map
 $\begin{array}{ccccc}
\Phi & : & \mathbb{Z}_{2}^{\gamma }\times \mathbb{Z}_{4}^{\delta } &
\rightarrow & \mathbb{Z}_{2}^{n}
\end{array}
$, where $n=\gamma +2\delta $ given by

\begin{center}
$\Phi (u,v)=(u,\phi (v_{1}),...,\phi (v_{\delta })),\forall u\in \mathbb{Z}
_{2}^{\gamma },\forall (v_{1},...,v_{\delta })\in \mathbb{Z}_{4}^{\delta }$
\end{center}

where
$\begin{array}{ccccc}
\phi & : & \mathbb{Z}_{4} & \rightarrow & \mathbb{Z}_{2}^{2}
\end{array}$ is the usual Gray map that is $\phi (0)=(0,0),\phi (1)=(0,1),\phi
(2)=(1,1),\phi (3)=(1,0)$. This Gray map is an isometry which transforms Lee
distances defined in $\mathbb{Z}_{2}^{\gamma }\times \mathbb{Z}_{4}^{\delta} $
 to Hamming distances defined in $\mathbb{Z}_{2}^{n}$, where $n=\gamma+2\delta $.

Let $v_{1}\in \mathbb{Z}_{2}^{\gamma }$ and $v_{2}\in \mathbb{Z}_{4}^{\delta
}$. Denote by $wt_{H}(v_{1})$ the Hamming weight of $v_{1}$ and $%
wt_{L}(v_{2})$ the Lee weight of $v_{2}$. For a vector $v=(v_{1},v_{2})\in
\mathbb{Z}_{2}^{\gamma }\times \mathbb{Z}_{4}^{\delta }$, define the weight
of $v$, denoted by $wt(v)$, as $wt_{H}(v_{1})+wt_{L}(v_{2})$, or
equivalently, the Hamming weight of $\Phi (v).$The Euclidean weight is given
by the relation $wt(v)=wt_{H}(v_{1})+wt_{E}(v_{2})$, where the Euclidean
weight $w_{E}\left( u\right)$ of a vector $u$ is $\sum\limits_{i=1}^{n}%
\min \left\{ u_{i},\left( 4-u_{i}\right) ^{2}\right\} $.
\begin{defi}
Let $C$ be a $\mathbb{Z}_{2}\mathbb{Z}_{4}-$ additive code, which is a sub
group of $\mathbb{Z}_{2}^{\gamma }\times \mathbb{Z}_{4}^{\delta }.$ We say
that the binary image $\Phi \left( C\right) $ is a $\mathbb{Z}_{2}\mathbb{Z
}_{4}-$ additive code of binary length $n=\gamma +2\delta $ and type $\left(
\gamma ,\delta ,\lambda ,\mu ;k\right) $, where $\lambda ,\mu $ and $k$ are
defined as above.
\end{defi}
Let $C$ be a $\mathbb{Z}_{2}\mathbb{Z}_{4}-$additive code. A $
\mathbb{Z}_{2}\mathbb{Z}_{4}-$additive code $C$ is not a free module, every
codeword is uniquely expressible in the form
\begin{center}
$\sum\limits_{i=1}^{\lambda }\lambda _{i}u_{i}+\sum\limits_{j=1}^{\mu }\mu
_{i}v_{j} $
\end{center}

where, $\lambda _{i}\in \mathbb{Z}_{2},$ $\mu _{j}\in \mathbb{Z}_{4}$ and for $%
\begin{array}{ccccc}
1 & \leq & i & \leq & \lambda%
\end{array}%
$ and $%
\begin{array}{ccccc}
1 & \leq & j & \leq & \mu%
\end{array}$,
$u_{i},v_{j}\in \mathbb{Z}_{2}^{\gamma }\times \mathbb{Z}_{4}^{\delta }$ of order two and order four,
respectively. Vectors $u_{i},v_{j}$ gives as a generator matrix $G$ of size $\left( \lambda+\mu\right) \times \left( \gamma+\delta\right)$ for the code $C$. Moreover, we can write $G$ as
\begin{center}
$G=\left[
\begin{tabular}{l|l}
$B_{1}$ & $2B_{3}$ \\ \hline
$B_{2}$ & $Q$
\end{tabular}
\right] $
\end{center}
where $B_{1}$, $B_{2}$ are matrices over $\mathbb{Z}$ of size $ \lambda\times \gamma$ and $\mu \times \gamma$, respectively; $B_{3}$ is a matrix over
$\mathbb{Z}_{4}$ of size $ \lambda \times \delta$ with all entries in $\left\lbrace 0, 1\right\rbrace  \subset \mathbb{Z}_{4}$; and $Q$ is a matrix over $\mathbb{Z}_{4}$ of size $\mu \times \delta$ with quaternary row vectors of order four.

In \cite{Borges}, it is shown that a $\mathbb{Z}_{2}\mathbb{Z}_{4}-$ additive code
is permutation equivalent to a $\mathbb{Z}_{2}\mathbb{Z}_{4}-$ additive code
with standard generator matrix of the form

\begin{equation}\label{equation:11}
G_{S}=\left[
\begin{tabular}{ll|lll}
$I_{k}$ & $T^{\prime }$ & $2T_{1}$ & $0$ & $0$ \\
$0$ & $0$ & $T_{2}$ & $2I_{\lambda -k}$ & $0$ \\ \hline
$0$ & $S^{\prime }$ & $S$ & $R$ & $I_{\lambda }$%
\end{tabular}%
\right] \tag{1}
\end{equation}

where $T^{\prime },T_{1},T_{2},R,S^{\prime }$ are a matrix over $\mathbb{Z}%
_{2}$ and $S$ is a matrix over $\mathbb{Z}_{4}$.

We define the inner product of vector $u,v\in \mathbb{Z}_{2}^{\gamma }\times
\mathbb{Z}_{4}^{\delta }$ as

\begin{center}
$\left\langle u,v\right\rangle _{\mathbb{Z}_{2}\mathbb{Z}_{4}}=2\left(
\sum\limits_{i=1}^{\gamma }u_{i}v_{i}\right) +\sum\limits_{j=\gamma
+1}^{\gamma +\delta }u_{j}v_{j}\in \mathbb{Z}_{4}$
\end{center}

The $\mathbb{Z}_{2}\mathbb{Z}_{4}-$ additive dual code of $C$, denoted by $
C^{\bot }$, is defined in the standard way

\begin{center}
$C^{\bot }=\left\{ v\in \mathbb{Z}_{2}^{\gamma }\times \mathbb{Z}_{4}^{\delta
}\left\vert \left\langle u,v\right\rangle _{\mathbb{Z}_{2}\mathbb{Z}_{4}}=0
\text{ for all }u\in C\right. \right\}$.
\end{center}
We will also call $C^{\bot }$ the additive dual code of $C$. We use the canonical generator matrix
as in \ref{equation:11} to construct a canonical generator matrix of $C^{\bot }$ , to get a parity-check matrix
\begin{equation}
H_{S}=\left[
\begin{tabular}{ll|lll}
$T^{\prime }$ & $I_{\gamma-k}$ & $0$ & $0$ & $2S^{\prime^{t}}$ \\
$0$ & $0$ & $0$ & $I_{\lambda -k}$ & $2R^{t}$ \\ \hline
$T_{1}^{t}$ & $0$ & $I_{\delta+k-\lambda-\mu}$ & $T_{2}^{t}$ & $-\left( S+RT_{2}\right)^{t} $%
\end{tabular}%
\right]
\end{equation}
where $T^{\prime },T_{1},T_{2},R,S^{\prime }$ are a matrix over $\mathbb{Z}%
_{2}$ and $S$ is a matrix over $\mathbb{Z}_{4}$.

\subsection{Covering Radius of Codes}
\bigskip In this section, we introduce the basic notions of the covering radius of
codes over $\mathbb{Z}_{2}\mathbb{Z}_{4}$.
The covering radius of a code $C$, denoted $r(C)$, is the smallest number $r$
such that the spheres covering radius of radius $r$ around the codewords of $
C$ cover the sets $\left( \mathbb{Z}_{2} \times \mathbb{Z}
_{4}\right)^{n} .$
The covering radius of a code $C$ over $\mathbb{Z}_{2}\mathbb{Z}_{4}$ with
respect to the Lee and Euclidien distances is given by

\begin{center}
$r_{L}(C)=\max_{u\in \left( \mathbb{Z}_{2}\times \mathbb{Z}_{4}\right) ^{n}
}\left\{ \min_{c\in C} d_{L}(u,c)\right\}$
\end{center}

and

\begin{center}
$r_{E}(C)=\max_{u\in \left( \mathbb{Z}_{2}\times \mathbb{Z}_{4}\right) ^{n}
}\left\{ \min_{c\in C} d_{E}(u,c)\right\}$
\end{center}

respectively. It is easy to see that $r_{L}(C)$, $r_{E}(C)$ are the minimum values $r_{L}$,
 $r_{E}$ such that

\begin{center}
$\left( \mathbb{Z}_{2} \times \mathbb{Z}_{4}\right)^{n} =\cup_{c \in C}
 S_{r_{L}}(c)$ and $\left( \mathbb{Z}_{2} \times \mathbb{Z}_{4}\right)^{n} =\cup_{c \in C}
S_{r_{E}}(c)$
\end{center}

respectively, where

\begin{center}
$S_{r_{L}}(u)=\left\{ v\in \left( \mathbb{Z}_{2} \times \mathbb{Z}
_{4}\right)^{n} \left \vert d(u,v)\leq r_{L} \right.\right\}$
\end{center}

and

\begin{center}
$S_{r_{E}}(u)=\left\{ v\in \left( \mathbb{Z}_{2} \times \mathbb{Z}
_{4}\right)^{n} \left\vert d(u,v)\leq r_{E} \right. \right\}$
\end{center}

for any element $u\in \left( \mathbb{Z}_{2} \times \mathbb{Z}
_{4}\right)^{n}.$

\begin{prop}\label{prop:1}
Let $C$ be a code over $\left( \mathbb{Z}_{2} \times \mathbb{Z}
_{4}\right)^{n}$ and $\Phi (C)$ the Gray map images of $C$.
Then $r_{L}\left( C\right) =r\left( \Phi (C)\right) .$
\end{prop}

The lower and upper bounds on the covering radius of codes over $\mathbb{Z}_{2}\mathbb{Z}
_{4}$ is given in several theorems and proposition (see, \cite{Gupta}, \cite{Manoj} and \cite{Pandian}).

\begin{prop}
For any code $C$ of length $n$ over $\mathbb{Z}_{2}\mathbb{Z}_{4}$,
\begin{center}
$\dfrac{2^{2n}}{\vert C\vert} \leqslant \sum_{i=0}^{r_{L}\left( C\right) }
\left( \begin{array}{c}
2n \\ i
 \end{array}\right) $
\end{center}
\begin{center}
$\dfrac{2^{2n}}{\vert C\vert} \leqslant \sum_{i=0}^{r_{E}\left( C\right) }
\left(V_{i}\right)$, where $\sum_{i=0}^{5n }
V_{i}x^{i}=\left( 1+3x+2x^{2}+x^{4}+x^{5}\right)^{n} $
\end{center}
\end{prop}
\pf
The proof of inequality over $\mathbb{Z}_{2}\mathbb{Z}_{4}$ is similar to the proof over $\mathbb{Z}_{4}$ given in \cite{Aoki}.

Let $C$ be a code over $\mathbb{Z}_{2}\mathbb{Z}_{4}$ and
\begin{center}
$s\left( C^{\bot}\right)=\vert \left\lbrace i \vert A_{i}\left( C^{\bot}\right)\neq 0,i\neq 0\right\rbrace \vert $
\end{center}
where $A_{i}\left( C^{\bot}\right)$ is the number of codewords of weight $i$ in $C^{\bot}$. Delsarte
in \cite{Dels} proved that the covering radius $r\left( C\right) $ of $C$ is given by
$r\left( C\right)\leqslant s\left( C^{\bot}\right)$. This
is known as the Delsarte bound.
Before to define the bound of  Delsarte we give the following Lemma
\begin{lem} \cite{Aoki} \label{lem:1}
For a code $C$ over $\mathbb{Z}_{2}\mathbb{Z}_{4}$, $r_{L}\left( C\right) \leq r_{E}\left( C\right)\leq 3r_{L}\left( C\right)$.
\end{lem}

\begin{thm}\label{thm:13}(Delesarte Bound)\cite{Aoki}
Let $C$ be a code over $\mathbb{Z}_{2}\mathbb{Z}_{4}$ then $r_{L}\left( C\right) \leq s\left( C^{\bot}\right) $ and $r_{E}\left( C\right) \leq 3s\left( C^{\bot}\right) $.

\end{thm}

A coset of the code $C$ defined by the vector $u\in \left( \mathbb{Z}_{2} \times \mathbb{Z}
_{4}\right)^{n}$ is the set

\begin{center}
$u+C=\left\{ u+v\left\vert v\in C\right. \right\}$
\end{center}

A coset leader of $C$ is a vector in $u+C$ of smallest weight. When the
code is linear its covering radius is equal to the weight of the heaviest
coset leader. Hence we have the following proposition

\begin{prop}\label{prop:2}\cite{Gupta}
The covering radius of the linear code $C$ is equal to the
maximum weight of a coset leader.
\end{prop}

The following result of Mattson is useful for computing covering
radii of codes over rings generalized easily from codes over finite fields.
\begin{prop}\label{prop:3}
If $C_{0}$ and $C_{1}$ are codes over $\mathbb{Z}_{2}\mathbb{Z}_{4}$ generated by
matrices $G_{0}$ and $G_{1}$ respectively and if $C$ is the code generated by

\begin{center}
$G=\left[
\begin{tabular}{l|l}
$0$ & $G_{1}$ \\ \hline
$G_{0}$ & $A$
\end{tabular}
\right]$
\end{center}

then $r_{d}(C)\leq r_{d}(C_{0})+r_{d}(C_{1})$ and the covering radius of $D$
(concatenation of $C_{0}$ and $C_{1}$) satisfy the following $r_{d}(D)\geq
r_{d}(C_{0})+r_{d}(C_{1})$ for all distances $d$ over $\mathbb{Z}_{2}\mathbb{Z}_{4}$.
\end{prop}
\section{The Covering Radius of $\mathbb{Z}_{2}\mathbb{Z}_{4}-$Repetition Codes}
\bigskip The repetition code $C$ over a finite field $\mathbb{F}_{q}=\left\{
\alpha _{0},\alpha _{1},\cdots ,\alpha _{q-2}\right\} $ is an $\left[ n,1,n
\right] -$ code such as $C=\left\{(\alpha \alpha\cdots \alpha )/\alpha \in \mathbb{F}%
_{q}\right\} $ . The covering radius of every $\left[ n,1,n\right] -$ code is $\lceil
\left( \frac{q-1}{q}\right)n\rceil $. In particular, this holds for the binary repetition code.
In \cite{Gupta}, various classes of repetition codes over $\mathbb{Z}_{4}$ have
been studied and their covering radius has been obtained. Now we
generalize those results for codes over $\mathbb{Z}_{2}\mathbb{Z}_{4}.$
Consider the repetition codes over $\mathbb{Z}_{2}\mathbb{Z}_{4}.$ One can
define seven basic repetition codes $C_{\alpha _{i}}$, $\left( 1\leq i\leq
7\right) $ of length $n$ over $\mathbb{Z}_{2}\mathbb{Z}_{4}$ generated by $%
G_{\alpha _{1}}=\left[ 0101\cdots 01\right] ,G_{\alpha _{2}}=\left[
0202\cdots 02\right] ,G_{\alpha _{3}}=\left[ 0303\cdots 03\right] ,G_{\alpha
_{4}}=\left[ 1010\cdots 10\right] ,G_{\alpha _{5}}=\left[ 1111\cdots 11%
\right] ,G_{\alpha _{6}}=\left[ 1212\cdots 12\right] ,G_{\alpha _{7}}=\left[
1313\cdots 13\right] .$
So the repetition codes are $C_{\alpha _{1}}=C_{\alpha _{3}}=\left\{ \left(
00\cdots 00\right) ,\left( 01\cdots 01\right) ,\left( 02\cdots
02\right) ,\left( 03\cdots 03\right) \right\} $, $C_{\alpha _{2}}=\\
\left\{
\left( 00\cdots 00\right) ,\left( 02\cdots 02\right) \right\} $, $%
C_{\alpha _{4}}=\left\{ \left( 00\cdots 00\right) ,\left( 10\cdots
10\right) \right\} $, $C_{\alpha _{5}}=C_{\alpha _{7}}=\\
\left\{ \left(
00\cdots 00\right) ,\left( 01\cdots 01\right) ,\left( 02\cdots
02\right) ,\left( 03\cdots 03\right) ,\left( 10\cdots 10\right) ,\left(
11\cdots 11\right) ,\left( 12\cdots 12\right) ,\left( 13\cdots
13\right) \right\} ,$ following $C_{\alpha _{6}}=\left\{ \left( 00\cdots
00\right) ,\left( 02\cdots 02\right) ,\left( 10\cdots 10\right) ,\left(
12\cdots 12\right) \right\} $. The theorems determine the covering radius
of $C_{\alpha _{i}}$ for $\left( 1\leq i\leq 7\right) .$
\begin{thm}\label{thm:1}
$r_{E}\left( C_{\alpha _{1}}\right) =r_{E}\left( C_{\alpha _{3}}\right) =
\frac{3n}{4}$ and $r_{L}\left( C_{\alpha _{1}}\right) =r_{L}\left( C_{\alpha
_{3}}\right) =\frac{3n}{2}.$
\end{thm}
\pf
We know that $r_{E}\left( C_{\alpha _{i}}\right) =\max_{x\in \left( \mathbb{Z
}_{2}\mathbb{Z}_{4}\right) ^{n}}\left\{ d_{E}\left( x,C_{\alpha _{i}}\right)
\right\} $. Let $x \in \left( \mathbb{Z}_{2}\mathbb{Z}_{4}\right)
^{n}.$ If $x$ has composition of
$\left( t_{0},t_{1},t_{2},t_{3},t_{4},t_{5},t_{6},t_{7}\right) $, where $
\sum\limits_{j=0}^{7}t_{i}=n$ then $d_{E}\left( x,\overline{00}\right)
= n-t_{0}+3t_{2}+t_{5}+4t_{6}+t_{7}$, $d_{E}\left( x,\overline{
01}\right) = n-t_{1}+3t_{3}+t_{6}+4t_{7}$, $d_{E}\left( x,
\overline{02}\right)= n-t_{2}+4t_{0}+t_{1}+3t_{4}+t_{7}$ and $
d_{E}\left( x,\overline{03}\right)= n-t_{3}+t_{0}+4t_{1}+t_{2}+3t_{5} $. Thus
 $d_{E}\left( x,C_{\alpha_{1}}\right) =$ min  $ ( n-t_{0}+3t_{2}+t_{5}+4t_{6}+t_{7})$
, $(n-t_{1}+3t_{3}+t_{6}+4t_{7})$ ,$( n-t_{2}+4t_{0}+t_{1}+3t_{4}+t_{7})$,
$(n-t_{3}+t_{0}+4t_{1}+t_{2}+3t_{5})$ $\leq \frac{3n}{4}$. If $ x=x_{1}x_{2}x_{3}x_{4}=
\overset{\frac{n}{4}}{\overbrace{00\cdots 00}}\overset{\frac{n}{4}}{
\overbrace{01\cdots 01}}\overset{\frac{n}{4}}{\overbrace{02\cdots 02}}
\overset{\frac{n}{4}}{\overbrace{03\cdots 03}}\in \left( \mathbb{Z}_{2}
\mathbb{Z}_{4}\right) ^{n}$, then $d_{E}\left( x,\overline{00}\right)
=d_{E}\left( x,\overline{02}\right) =d_{E}\left( x,\overline{03}\right) =
\frac{n}{8}+4\left( \frac{n}{8}\right) +\frac{n}{8}=\frac{3n}{4}.$ Thus $
r_{E}\left( C_{\alpha _{1}}\right) \geq \frac{3n}{4}.$ Hence $r_{E}\left(
C_{\alpha _{1}}\right) =r_{E}\left( C_{\alpha _{3}}\right) =\frac{3n}{4}.$
The gray map $\Phi (C_{\alpha _{1}})$ is a binary repetition code of length
$3n$ hence $r_{L}\left( C_{\alpha _{1}}\right) =r_{L}\left( C_{\alpha
_{3}}\right) =\frac{3n}{2}.$
\qed

\begin{thm}\label{thm:2}
$r_{E}\left( C_{\alpha _{5}}\right) =r_{E}\left( C_{\alpha _{7}}\right) =n$
and $r_{L}\left( C_{\alpha _{5}}\right) =r_{L}\left( C_{\alpha _{7}}\right) =%
\frac{3n}{2}.$
\end{thm}
\pf

See the first part of Theorem \ref{thm:1} is to that $r_{E}\left(
C_{\alpha _{5}}\right) \leq n.$ If\\
 $x=\overset{\frac{n}{8}}{\overbrace{%
00\cdots 00}}\overset{\frac{n}{8}}{\overbrace{01\cdots 01}}\overset{\frac{n}{%
8}}{\overbrace{02\cdots 02}}\overset{\frac{n}{8}}{\overbrace{03\cdots 03}}%
\overset{\frac{n}{8}}{\overbrace{10\cdots 10}}\overset{\frac{n}{8}}{%
\overbrace{11\cdots 11}}\overset{\frac{n}{8}}{\overbrace{12\cdots 12}}%
\overset{\frac{n}{8}}{\overbrace{13\cdots 13}}\in \left( \mathbb{Z}_{2}%
\mathbb{Z}_{4}\right) ^{n}$, then\\
 $d_{E}\left( x,\overline{00}\right)
=d_{E}\left( x,\overline{01}\right) =d_{E}\left( x,\overline{02}\right)
=d_{E}\left( x,\overline{03}\right) =d_{E}\left( x,\overline{10}\right)
=d_{E}\left( x,\overline{11}\right) =d_{E}\left( x,\overline{12}\right)
=d_{E}\left( x,\overline{13}\right) =\frac{n}{16}+4\left( \frac{n}{16}%
\right) +\frac{n}{16}+\frac{n}{16}+\frac{n}{8}+\frac{n}{16}+4\left( \frac{n}{%
16}\right) +\frac{n}{8}=n$. Thus $r_{E}\left( C_{\alpha _{5}}\right) \geq n.$ Hence $r_{E}\left(
C_{\alpha _{5}}\right) =r_{E}\left( C_{\alpha _{7}}\right) =n.$
\qed

\begin{thm}\label{thm:3}
$r_{E}\left( C_{\alpha _{2}}\right) =n,r_{E}\left( C_{\alpha _{4}}\right) =%
\frac{n}{4},r_{E}\left( C_{\alpha _{6}}\right) =\frac{5n}{4}$ and $%
r_{L}\left( C_{\alpha _{2}}\right) =r_{L}\left( C_{\alpha _{4}}\right)
=r_{L}\left( C_{\alpha _{6}}\right) =\frac{3n}{2}.$
\end{thm}
\pf
The proof is similar to proof of Theorem \ref{thm:1} and \ref{thm:2}, hence omitted.
\qed

 In order to determine the covering radius of Simplex code of type $\alpha $ and $\beta$
 over $\mathbb{Z}_{2}\mathbb{Z}_{4}$, we have to define a
block repetition code over $\mathbb{Z}_{2}\mathbb{Z}_{4}$ and find its
covering radius.
Thus the covering radius of the block repetition code $B{Rep}^{n}:(n=n_{1}+n_
{2}+n_{3}+n_{4}+n_{5}+n_{6}+n_{7},2^{3},d_{L}=6n ,d_{E}=\min
\{(n_{1}+4n_{2}+n_{3}+n_{4}+2n_{5}+5n_{6}+2n_{7}),$
$\left( n_{1}+4n_{2}+n_{3}+n_{5}+4n_{6}+n_{7}\right) ,\\
\left(4n_{1}+n_{2}+4n_{3}+4n_{5}+n_{6}+4n_{7}\right) ,\left(
n_{4}+n_{5}+n_{6}+n_{7}\right) ,
\left(
4n_{1}+4n_{3}+n_{4}+5n_{5}+n_{6}+5n_{7}\right) \}$ generated by
\begin{equation*}
G=\left( \overset{n_{1}}{\overbrace{01\cdots 01}}\overset{n_{2}}{\overbrace{02\cdots
02}}\overset{n_{3}}{\overbrace{03\cdots 03}}\overset{n_{4}}{\overbrace{10\cdots 10}}%
\overset{n_{5}}{\overbrace{11\cdots 11}}\overset{n_{6}}{\overbrace{12\cdots 12}}%
\overset{n_{7}}{\overbrace{13\cdots 13}}\right)
\end{equation*}

is given in the following theorems.

\begin{thm}\label{thm:4}
$r_{E}\left( B{Re}p
^{n_{1}+n_{2}+n_{3}+n_{4}+n_{5}+n_{6}+n_{7}}\right) =\frac{1}{4}\left[
3\left( n_{1}+n_{3}\right) +n_{4}+5n_{6}\right] +\left(
n_{2}+n_{5}+n_{7}\right) $ and $r_{E}\left( B{Re}p_{\alpha
}^{7n}\right) =6n$
\end{thm}

\pf
By proposition \ref{prop:3} and Theorem \ref{thm:1}, \ref{thm:2} and \ref{thm:3} we have\\
$r_{E}\left( B{Re}p^{n_{1}+n_{2}+n_{3}+n_{4}+n_{5}+n_{6}+n_{7}}
\right) \geq \frac{1}{4}\left[ 3\left(
n_{1}+n_{3}\right) +n_{4}+5n_{6}\right] +\left( n_{2}+n_{5}+n_{7}\right) $.
Let\\
$x=x_{1}x_{2}x_{3}x_{4}x_{5}x_{6}x_{7}\in \left( \mathbb{Z}_{2}\mathbb{Z}
_{4}\right) ^{n_{1}+n_{2}+n_{3}+n_{4}+n_{5}+n_{6}+n_{7}}$ with $%
x_{1},x_{2},x_{3},x_{4},x_{5},x_{6},x_{7}$ have compositions of $\left(
a_{0},a_{1},a_{2},a_{3},a_{4},a_{5},a_{6},a_{7}\right), \left(b_{0},b_{1},b_{2},b_{3},b_{4},b_{5},b_{6},b_{7}\right), \left(
c_{0},c_{1},c_{2},c_{3},c_{4},c_{5},c_{6},c_{7}\right),\\
\left(
d_{0},d_{1},d_{2},d_{3},d_{4},d_{5},d_{6},d_{7}\right)$,
$\left( e_{0},e_{1},e_{2},e_{3},e_{4},e_{5},e_{6},e_{7}\right) ,\left(
f_{0},f_{1},f_{2},f_{3},f_{4},f_{5},f_{6},f_{7}\right), \\
\left(
g_{0},g_{1},g_{2},g_{3},g_{4},g_{5},g_{6},g_{7}\right) $ such that
 $
n_{1}=\sum\limits_{j=0}^{7}a_{i},n_{2}=\sum\limits_{j=0}^{7}b_{i},n_{3}=
\sum\limits_{j=0}^{7}c_{i},n_{4}=\sum\limits_{j=0}^{7}d_{i},n_{5}=\sum
\limits_{j=0}^{7}e_{i},n_{6}=\sum\limits_{j=0}^{7}f_{i},n_{7}=\sum
\limits_{j=0}^{7}g_{i}.$

$d_{E}\left( x,\overline{00}\right) =
n_{1}-a_{0}+3a_{2}+a_{5}+4a_{6}+a_{7} +
n_{2}-b_{0}+3b_{2}+b_{5}+4b_{6}+b_{7} +
n_{3}-c_{0}+3c_{2}
+c_{5}+4c_{6}+c_{7} +
n_{4}-d_{0}+3d_{2}+d_{5}+4d_{6}+d_{7} +
n_{5}-e_{0}+3e_{2}+e_{5}+4e_{6}+e_{7} +
n_{6}-f_{0}+3f+f_{5}+4f_{6}+f_{7} +
n_{7}-g_{0}+3g_{2}+g_{5}+4g_{6}+g_{7} $,
where $\overline{00}=0000\cdots 00$ is the first vector of $B{Re}
p^{n_{1}+n_{2}+n_{3}+n_{4}+n_{5}+n_{6}+n_{7}}.$

$d_{E}\left( x,\overline{y_{1}}\right) =
n_{1}-a_{1}+3a_{3}+a_{6}+4a_{7} +
n_{2}-b_{2}+4b_{0}+b_{1}+3b_{4}+b_{7} +
n_{3}-c_{3}+c_{0}+4c_{1}+c_{2}+3c_{5} +
n_{4}-d_{4}+d_{1}+4d_{2}+d_{3}+3d_{6} +
n_{5}-e_{5}+e_{2}+4e_{3}+3e_{7} +
n_{6}-f_{6}+3f_{0}+f_{3}+4f_{4}+f_{5} +
n_{7}-g_{7}+3g_{1}+g_{4}+4g_{5}+g_{6} $,
where $\overline{y_{1}}=\overset{n_{1}}{\overbrace{01\cdots 01}}\overset{
n_{2}}{\overbrace{02\cdots 02}}\overset{n_{3}}{\overbrace{03\cdots 03}}
\overset{n_{4}}{\overbrace{10\cdots 10}}\overset{n_{5}}{\overbrace{11\cdots
11}}\overset{n_{6}}{\overbrace{12\cdots 12}}\overset{n_{7}}{\overbrace{
13\cdots 13}}$ is the second vector of $B{Re}p
^{n_{1}+n_{2}+n_{3}+n_{4}+n_{5}+n_{6}+n_{7}}.$

$d_{E}\left( x,\overline{y_{2}}\right) =
n_{1}-a_{1}+3a_{3}+a_{6}+4a_{7} +
n_{2}-b_{2}+4b_{0}+b_{1}+3b_{4}+b_{7} +
n_{3}-c_{3}+c_{0}+4c_{1}+c_{2}+3c_{5} +
n_{4}-d_{0}+3d_{2}+d_{5}+4d_{6}+d_{7}+
n_{5}-e_{1}+3e_{3}+e_{6}+4e_{7}+
n_{6}-f_{2}+4f_{0}+f_{1}+3f_{4}+f_{7}+
n_{7}-g_{3}+g_{0}+4g_{1}+g_{2}+3g_{5} $,
where $\overline{y_{2}}=\overset{n_{1}}{\overbrace{01\cdots 01}}\overset{
n_{2}}{\overbrace{02\cdots 02}}\overset{n_{3}}{\overbrace{03\cdots 03}}
\overset{n_{4}}{\overbrace{00\cdots 00}}\overset{n_{5}}{\overbrace{01\cdots
01}}\overset{n_{6}}{\overbrace{02\cdots 02}}\overset{n_{7}}{\overbrace{
03\cdots 03}}$ is the third vector of $B{Re}p^{n_{1}+n_{2}+n_{3}+n_{4}+n_{5}+n_{6}+n_{7}}.$

$d_{E}\left( x,\overline{y_{3}}\right) =
n_{1}-a_{2}+4a_{0}+a_{1}+3a_{4}+a_{7} +
n_{2}-b_{1}+3b_{3}+b_{6}+4b_{7} +
n_{3}-c_{2}+4c_{0}+c_{1}+3c_{4}+c_{7}+
n_{4}-d_{0}+3d_{2}+d_{5}+4d_{6}+d_{7}+
n_{5}-e_{2}+4e_{0}+e_{1}+3e_{4}+e_{7}+
n_{6}-f_{1}+3f_{3}+f_{6}+4f_{7}+
n_{7}-g_{2}+4g_{0}+g_{1}+3g_{4}+g_{7}$,
where $\overline{y_{3}}=\overset{n_{1}}{\overbrace{02\cdots 02}}\overset{
n_{2}}{\overbrace{01\cdots 01}}\overset{n_{3}}{\overbrace{02\cdots 02}}
\overset{n_{4}}{\overbrace{00\cdots 00}}\overset{n_{5}}{\overbrace{02\cdots
02}}\overset{n_{6}}{\overbrace{01\cdots 01}}\overset{n_{7}}{\overbrace{
02\cdots 02}}$ is the fourth vector of $B{Re}p
^{n_{1}+n_{2}+n_{3}+n_{4}+n_{5}+n_{6}+n_{7}}.$

$d_{E}\left( x,\overline{y_{4}}\right) =
n_{1}-a_{3}+a_{0}+4a_{1}+a_{2}+3a_{5}+
n_{2}-b_{2}+4b_{0}+b_{1}+3b_{4}+b_{7}+
n_{3}-c_{1}+3c_{3}+c_{6}+4c_{7}+
n_{4}-d_{0}+3d_{2}+d_{5}+4d_{6}+d_{7}+
n_{5}-e_{3}+e_{0}+4e_{1}+e_{2}+3e_{5}+
n_{6}-f_{2}+4f_{0}+f_{1}+3f_{4}+f_{7}+
n_{7}-g_{1}+3g_{3}+g_{6}+4g_{7}+ $,
where $\overline{y_{4}}=\overset{n_{1}}{\overbrace{03\cdots 03}}\overset{
n_{2}}{\overbrace{02\cdots 02}}\overset{n_{3}}{\overbrace{01\cdots 01}}
\overset{n_{4}}{\overbrace{00\cdots 00}}\overset{n_{5}}{\overbrace{03\cdots
03}}\overset{n_{6}}{\overbrace{10\cdots 10}}\overset{n_{7}}{\overbrace{
01\cdots 01}}$ is the fifth vector of $B{Re}p
^{n_{1}+n_{2}+n_{3}+n_{4}+n_{5}+n_{6}+n_{7}}.$

$d_{E}\left( x,\overline{y_{5}}\right) =
n_{1}-a_{0}+3a_{2}+a_{5}+4a_{6}+a_{7}+
n_{2}-b_{0}+3b_{2}+b_{5}+4b_{6}+b_{7}+
n_{3}-c_{0}+3c_{2}+c_{5}+4c_{6}+c_{7}+
n_{4}-d_{4}+d_{1}+4d_{2}+d_{3}+3d_{6}+
n_{5}-e_{4}+e_{1}+4e_{2}+e_{3}+3e_{6}+
n_{6}-f_{4}+d_{f1}+4f_{2}+f_{3}+3f_{6}+
n_{4}-g_{4}+g_{1}+4g_{2}+g_{3}+3g_{6} $,
where $\overline{y_{5}}=\overset{n_{1}}{\overbrace{00\cdots 00}}\overset{
n_{2}}{\overbrace{00\cdots 00}}\overset{n_{3}}{\overbrace{00\cdots 00}}
\overset{n_{4}}{\overbrace{10\cdots 10}}\overset{n_{5}}{\overbrace{10\cdots
10}}\overset{n_{6}}{\overbrace{10\cdots 10}}\overset{n_{7}}{\overbrace{
10\cdots 10}}$ is the sixth vector of $B{Re}p
^{n_{1}+n_{2}+n_{3}+n_{4}+n_{5}+n_{6}+n_{7}}.$

$d_{E}\left( x,\overline{y_{6}}\right) =
n_{1}-a_{2}+4a_{0}+a_{1}+3a_{4}+a_{7}+
n_{2}-b_{0}+3b_{2}+b_{5}+4b_{6}+b_{7}+
n_{3}-c_{2}+4c_{0}+c_{1}+3c_{4}+c_{7}+
n_{4}-d_{4}+d_{1}+4d_{2}+d_{3}+3d_{6}+
n_{6}-e_{6}+3e_{0}+e_{3}+4e_{4}+e_{5}+
n_{6}-f_{4}+f_{1}+4f_{2}+f_{3}+3f_{6}+
n_{7}-g_{6}+3g_{0}+g_{3}+4g_{4}+g_{5}+ $,
where $\overline{y_{6}}=\overset{n_{1}}{\overbrace{02\cdots 02}}\overset{
n_{2}}{\overbrace{00\cdots 00}}\overset{n_{3}}{\overbrace{02\cdots 02}}
\overset{n_{4}}{\overbrace{10\cdots 10}}\overset{n_{5}}{\overbrace{12\cdots
12}}\overset{n_{6}}{\overbrace{10\cdots 10}}\overset{n_{7}}{\overbrace{
12\cdots 12}}$ is the seventh vector of $B{Re}p
^{n_{1}+n_{2}+n_{3}+n_{4}+n_{5}+n_{6}+n_{7}}.$

$d_{E}\left( x,\overline{y_{7}}\right) =
n_{1}-a_{3}+a_{0}+4a_{1}+a_{2}+3a_{5}+
n_{2}-b_{2}+4b_{0}+b_{1}+3b_{4}+b_{7}+
n_{3}-c_{1}+3c_{3}+c_{6}+4c_{7}+
n_{4}-d_{4}+d_{1}+4d_{2}+d_{3}+3d_{6}+
n_{7}-e_{3}+e_{0}+4e_{1}+e_{2}+3e_{5}+
n_{2}-f_{2}+4f_{0}+f_{1}+3f_{4}+f_{7}+
n_{5}-g_{5}+g_{2}+4g_{3}+3g_{7} $,
where $\overline{y_{7}}=\overset{n_{1}}{\overbrace{03\cdots 02}}\overset{
n_{2}}{\overbrace{302\cdots 02}}\overset{n_{3}}{\overbrace{01\cdots 01}}
\overset{n_{4}}{\overbrace{10\cdots 10}}\overset{n_{5}}{\overbrace{13\cdots
13}}\overset{n_{6}}{\overbrace{12\cdots 12}}\overset{n_{7}}{\overbrace{
11\cdots 11}}$ is the eighth vector of $B{Re}p_{\alpha
}^{n_{1}+n_{2}+n_{3}+n_{4}+n_{5}+n_{6}+n_{7}}.$ Thus $r_{E}\left( B{Re}
p^{n_{1}+n_{2}+n_{3}+n_{4}+n_{5}+n_{6}+n_{7}}\right) \leq \frac{1}{
4}\left[ 3\left( n_{1}+n_{3}\right) +n_{4}+5n_{6}\right] +\left(
n_{2}+n_{5}+n_{7}\right) .$

The code has constant Lee weight $6n$. Thus for $x=1111\cdots 11\in \left(
\mathbb{Z}_{2}\mathbb{Z}_{4}\right) ^{7n}$, we have $d_{L}\left( x, B{Re}
p^{7n}\right) =6n$.
\qed

\section{$\mathbb{Z}_{2}\mathbb{Z}_{4}$-Simplex Code of Type $\alpha $ and $\beta $}
\bigskip In this part, we have going to study the simplex code over $\mathbb{
Z}_{2}$ $\mathbb{Z}_{4}$ of type $\alpha $ and $\beta$ we will discuss the properties of these codes.
Type $\alpha $ simplex code $S_{k}^{\alpha }$ is linear code over $\mathbb{Z}
_{2}$ $\mathbb{Z}_{4}$ with parameters $\left[ 2^{3k+1},2k,d_{L},d_{E}\right] $ has a generator matrix which after a suitable permutation of coordinates can be written in the form
\begin{equation}\label{equation:1}
\Theta _{k}^{\alpha }=\left[
\begin{tabular}{l|l|l|l|l|l|l|l}
$m_{k}^{\alpha }$ & $m_{k}^{\alpha }$ & $\cdots $ & $m_{k}^{\alpha }$ & $
G_{k}^{\alpha }$ & $G_{k}^{\alpha }$ & $\cdots $ & $G_{k}^{\alpha }$
\end{tabular}%
\right] ,\text{ for }k\geq 1  \tag{1}
\protect\end{equation}
where $m_{k}^{\alpha }$ is a generator matrix of binary simplex code $
S_{2,k}^{\alpha }$ of type $\alpha $ repeat $2^{2k}$ times in $\Theta
_{k}^{\alpha }$ and $G_{k}^{\alpha }$ is a generator matrix of simplex code $
S_{4,k}^{\alpha }$ over $\mathbb{Z}_{4}$ of type $\alpha $ repeat $2^{k}$
times in $\Theta _{k}^{\alpha }$

Type $\beta $ simplex code $S_{k}^{\beta }$ is a punctured version of $
S_{k}^{\alpha }$ with the parameters \\
$\left[2^{3(k-1)}(2^{k}-1),2k,d_{L},d_{E}\right] $ and has a generator matrix $
\Theta _{k}^{\beta }$ of the form
\begin{equation}\label{equation:2}
\Theta _{k}^{\beta }=\left[
\begin{tabular}{l|l|l|l|l|l|l}
$m_{k}^{\beta }$ & $m_{k}^{\beta }$ & $\cdots $ & $m_{k}^{\beta }$ & $
G_{k}^{\beta }$ & $\cdots $ & $G_{k}^{\beta }$%
\end{tabular}%
\right] ,\text{ for }k\geq 3  \tag{2}
\end{equation}
where $m_{k}^{\beta }$ is a generator matrix of binary simplex code of type $
\beta $ repeat $2^{k}$ times in $\Theta _{k}^{\beta }$ and $G_{k}^{\beta }$
is a generator matrix of simplex code over $\mathbb{Z}_{4}$ of type $\beta $
repeat $2^{k-1}$ times in $\Theta _{k}^{\beta }.$

\subsection{The Covering Radius of $\mathbb{Z}_{2}\mathbb{Z}_{4}$-Simplex Codes of Type $\alpha$ and $\beta$}
The following two results are two upper bounds of the covering radius of codes over $\mathbb{Z}_{2}\mathbb{Z}_{4}$ with respect to Lee and Euclidean weight.
\begin{thm}\label{thm:5}
$r_{L}(S_{k}^{\alpha })=2^{3k+1}$ and $r_{E}(S_{k}^{\alpha })\leq 2^{k}\cdot
\left( \frac{17\cdot 2^{2k}-2}{6}\right) .$
\end{thm}
\pf
$\mathbb{Z}_{2}\mathbb{Z}_{4}$-Simplex code of type $\alpha $ is of constant
Lee weight equal $2^{3k+1}$. Hence by definition,$r_{L}(S_{k}^{\alpha })\geq
2^{3k+1}.$ On the other hand by the matrix \ref{equation:1}, the result of
Mastton (see Proposition \ref{prop:3}) for finite rings and Theorem \ref{thm:4}, we get

\begin{center}
$\begin{array}{ccc}
r_{L}(S_{k}^{\alpha }) & \leq & r_{L}(2^{2k}S_{2,k}^{\alpha
})+r_{L}(2^{k}S_{4,k}^{\alpha }) \\
& \leq & 2^{2k}r_{L}(S_{2,k}^{\alpha })+2^{k}r_{L}(S_{4,k}^{\alpha }) \\
& \leq & 2^{2k}r_{H}(S_{2,k}^{\alpha })+2^{k}r_{L}(S_{4,k}^{\alpha }) \\
& \leq & 2^{2k}\left[ \left( 2^{k}+2^{k-1}+\cdots +2^{1}\right)
+r_{L}(S_{2,1}^{\alpha })\right] \\
&  & +2^{k}\left[ \left( 3\cdot 2^{2(k-1)}+3\cdot 2^{2(k-2)}+\cdots +3\cdot
2^{2\cdot 1}\right) +r_{L}(S_{4,1}^{\alpha })\right] \\
& \leq & 2^{2k}\left[ \left( 2^{k}-1\right) +1\right] +2^{k}\left[ \left(
2^{2k}-2\right) +1\right] \\
& \leq & 2^{2k}\cdot 2^{k}+2^{k}\cdot 2^{2k} \\
&  & 2^{3k+1}
\end{array}$
\end{center}

Thus $r_{L}(S_{k}^{\alpha })=2^{3k+1}$.

Similar arguments can be used to show that (using Theorem \ref{thm:4})

\begin{center}
$\begin{array}{ccc}
r_{E}(S_{k}^{\alpha }) & \leq & r_{E}(2^{2k}S_{2,k}^{\alpha
})+r_{E}(2^{k}S_{4,k}^{\alpha }) \\
& \leq & 2^{2k}r_{E}(S_{2,k}^{\alpha })+2^{k}r_{E}(S_{4,k}^{\alpha }) \\
& \leq & 2^{2k}r_{H}(S_{2,k}^{\alpha })+2^{k}r_{E}(S_{4,k}^{\alpha }) \\
& \leq & 2^{2k}\cdot 2^{k}+2^{k}\cdot \frac{11\left( 2^{2k}-1\right) +9}{6}
\\
& \leq & 2^{k}\cdot \left( \frac{17\cdot 2^{2k}-2}{6}\right)
\end{array}$.
\end{center}
\qed

Similar arguments will compute the covering radius of $\mathbb{Z}_{2}\mathbb{
Z}_{4}$-Simplex code of type $\beta .$

\begin{thm}\label{thm:6}
The covering radius of $\mathbb{Z}_{2}\mathbb{Z}_{4}$-Simplex code of type $\beta $ is given by
\begin{enumerate}
\item $r_{L}(S_{k}^{\beta })\leq 2^{2k}\left( 2^{k}-1\right) +2^{k}\left(
2^{k-1}-2\right) $

\item $r_{E}(S_{k}^{\beta })\leq 2^{k}\left( \frac{17}{6}\cdot 2^{2k}-2\cdot
2^{k}-\frac{443}{6}\right) $
\end{enumerate}
\end{thm}
\pf
By the matrix \ref{equation:2}, proposition \ref{prop:3} and theorem \ref{thm:4}, we get

\begin{center}
$\begin{array}{ccc}
r_{L}(S_{k}^{\beta }) & \leq & r_{L}(2^{2k}S_{2,k}^{\beta
})+r_{L}(2^{k}S_{4,k}^{\beta }) \\
& \leq & 2^{2k}r_{L}(S_{2,k}^{\beta })+2^{k}r_{L}(S_{4,k}^{\beta }) \\
& \leq & 2^{2k}r_{H}(S_{2,k}^{\beta })+2^{k}r_{L}(S_{4,k}^{\beta }) \\
& \leq & 2^{2k}\left( 2^{k-1}-1\right) +2^{k}\left( 2^{k-1}\left(
2^{k}-1\right) -2\right) \\
&  & 2^{2k}\left( 2^{k}-1\right) +2^{k}\left( 2^{k-1}-2\right)
\end{array}$
\end{center}

Similar arguments can be used to show that (using Theorem \ref{thm:4})

\begin{center}
$\begin{array}{ccc}
r_{E}(S_{k}^{\beta }) & \leq & r_{E}(2^{2k}S_{2,k}^{\beta
})+r_{E}(2^{k}S_{4,k}^{\beta }) \\
& \leq & 2^{2k}r_{E}(S_{2,k}^{\beta })+2^{k}r_{E}(S_{4,k}^{\beta }) \\
& \leq & 2^{2k}r_{H}(S_{2,k}^{\beta })+2^{k}r_{E}(S_{4,k}^{\beta }) \\
& \leq & 2^{2k}\left( 2^{k-1}-1\right) +2^{k}\left[ 2^{k}\left(
2^{k+1}-1\right) +\frac{1}{3}\left( 4^{k}-1\right) -\frac{147}{2}\right] \\
& \leq & 2^{k}\left( \frac{17}{6}\cdot 2^{2k}-2\cdot 2^{k}-\frac{443}{6}
\right)
\end{array}$
\end{center}
\qed
\begin{thm}\label{thm:14}
$r_{L}(S_{k}^{\alpha^{\bot}})=r_{L}(S_{k}^{\beta ^{\bot}})=1$, $r_{E}(S_{k}^{\alpha^{\bot}})\leq 3$ and $r_{L}(S_{k}^{\beta ^{\bot}})\leq 3$
\end{thm}
\pf
The bound of Delsarte gives, $r_{L}(S_{k}^{\alpha^{\bot}})\leq 1$ and $r_{L}(S_{k}^{\beta ^{\bot}})\leq 1$. So equality is satisfied. Lemma \ref{lem:1} show that $r_{E}(S_{k}^{\alpha^{\bot}})\leq 3$ and $r_{L}(S_{k}^{\beta ^{\bot}})\leq 3$.
\qed
\section{$\mathbb{Z}_{2}\mathbb{Z}_{4}$-MacDonald Code of Type $\alpha $ and $ \beta$ and their Covering Radius}
\bigskip The MacDonald code $\mathcal{M}_{k,u}(q)$ over the finite field $
\mathbb{F}_{q}$ is a unique $\left[ \frac{q^{k}-q^{u}}{q-1},k,q^{k-1}-q^{u-1}
\right] $ code in which every nonzero codeword has weight either $q^{k-1}$
or $q^{k-1}-q^{u-1}$. Let $ 1 \leq u \leq k-1 $, $m_{k,u}^{\alpha }$(resp.,$ m_{k,u}^{\beta }$) be the matrix obtained from the matrix of binary simplex code $m_{k}^{\alpha }$(resp.,
$m_{k}^{\beta }$) by deleting columns corresponding to the columns of the matrix $
m_{u}^{\alpha }$ and $0_{2^{2^{u}}\times\left( k-u \right )}$(resp., $ m_{u}^{\beta }$
and $0_{2^{2^{u}}\times \left( k-u \right )}$). i.e,
\begin{equation}
m_{k,u}^{\alpha }=\left[
\begin{array}{ccc}
m_{k}^{\alpha } & \left\backslash {}\right. & \frac{0_{2^{2^{u}}\times
\left( k-u\right)}}{m_{u}^{\alpha }}
\end{array}
\right] \tag {3}
\end{equation}
and

for $k\geq3$
\begin{equation}
m_{k,u}^{\beta }=\left[
\begin{array}{ccc}
m_{k}^{\beta} & \left\backslash {}\right. & \frac{0_{2^{2^{u}}\times
\left( k-u\right)}}{m_{u}^{\beta}}
\end{array}
\right]\tag {4}
\end{equation}
In \cite{Colbourn} authors have defined the MacDonald codes over $
\mathbb{Z}_{4}$ using the generator matrices of simplex codes. For $ 1\leq u \leq k-1 $,
let $G_{k,u}^{\alpha }$ (resp.,  $G_{k,u}^{\beta }$) be the matrix
obtained from $G_{k}^{\alpha }$ (resp., $ G_{k}^{\beta }$) by deleting
columns corresponding to the columns of the matrix $G_{u}^{\alpha }$ and $0_{2^{2^{u}}\times
\left( k-u \right )}$ $(resp., G_{_{u}}^{\beta }$ and $0_{2^{2^{u}}\times
\left( k-u \right )}) $ i.e,
\begin{equation}
G_{k,u}^{\alpha}=\left[
\begin{array}{ccc}
G_{k}^{\alpha} & \left\backslash {}\right. & \frac{0_{2^{2^{u}}\times
\left( k-u \right )}}{G_{u}^{\alpha}}
\end{array}
\right]  \tag{5}
\end{equation}
and

for $k\geq 3$
\begin{equation}\label{equation:4}
G_{k,u}^{\beta }=\left[
\begin{array}{ccc}
G_{k}^{\beta } & \left\backslash {}\right. & \frac{0_{2^{2^{u}}\times
\left( k-u \right )}}{G_{u}^{\beta }}
\end{array}
\right]  \tag{6}
\end{equation}
Now, we will construct the MacDonald codes over $\mathbb{Z}_{2}\mathbb{Z}
_{4} $ of type $\alpha $ and $\beta $ by using the generator matrix of the $
\mathbb{Z}_{2}\mathbb{Z}_{4}-$simplex codes of type $\alpha $ and $\beta $.
If $ 1 \leq u \leq k-1 $, let $\Theta _{k,u}^{\alpha }$( resp., $\Theta _{k,u}^{\beta }$) be the
matrix of MacDonald codes $\mathcal{M}_{k,u}^{\alpha }$ (resp.,$\mathcal{M}_{k,u}^{\beta }$) with parametrs $\left[ 2^{3k+1}-2^{k+u}\left( 2^{k}-2^{u}\right) \right]$ (resp.,
$\left[2^{2k-1}\left(2^{2k-1}+1 \right)\left( 2^{k}-1\right)-2^{k+u-1}\left
(2^{2u-3}+1\right)\left(2^{u}-1 \right)\right]$) obtained from $\Theta _{k}^{\alpha }$
(resp., $\Theta _{k}^{\beta
}$) by deleting columns corresponding to the columns of the matrix $\Theta
_{u}^{\alpha }$ and ${0_{2^{2^{u}}\times
\left( k-u \right )}} $ (resp., $\Theta _{u}^{\beta }$ and ${0_{2^{2^{u}}\times
\left( k-u \right )}}$). i.e,

for $k\geq 1$
\begin{equation}\label{equation:5}
\Theta _{k,u}^{\alpha }=\left[
\begin{tabular}{l|l|l|l|l|l}
$
\begin{array}{ccc}
m_{k,u}^{\alpha }
\end{array}
$ & $\cdots $ & $
\begin{array}{ccc}
m_{k,u}^{\alpha }
\end{array}
$ & $
\begin{array}{ccc}
G_{k,u}^{\alpha }
\end{array}
$ & $\cdots $ & $
\begin{array}{ccc}
G_{k,u}^{\alpha}
\end{array}
$
\end{tabular}
\right] ,  \tag{7}
\end{equation}
where $m_{k,u}^{\alpha }$ (resp., $G_{k,u}^{\alpha }$) repeat $2^{2k}$ (resp., $2^{k}$)times in $\Theta _{k,u}^{\alpha }$.

and for $k\geq 3$
\begin{equation}\label{equation:6}
\Theta _{k,u}^{\beta }=\left[
\begin{tabular}{l|l|l|l|l|l}
$
\begin{array}{ccc}
m_{k,u}^{\beta}
\end{array}
$ & $\cdots $ & $
\begin{array}{ccc}
m_{k,u}^{\beta }
\end{array}
$ & $
\begin{array}{ccc}
G_{k,u}^{\beta }
\end{array}
$ & $\cdots $ & $
\begin{array}{ccc}
G_{k,u}^{\beta }
\end{array}
$
\end{tabular}
\right] ,  \tag{8}
\end{equation}
where $m_{k,u}^{\beta}$ (resp., $G_{k,u}^{\beta }$) repeat $2^{k}$ (resp., $2^{k-1}$)times in $\Theta _{k,u}^{\beta }$.

Next theorems provides basic bounds on the covering radius of MacDonald
codes over $\mathbb{Z}_{2}\mathbb{Z}_{4}$ of type $\alpha $ .

\begin{thm}
for $
\begin{array}{ccccc}
u & \leq & r & \leq & k
\end{array}
$
\begin{enumerate}
\item $
\begin{array}{ccc}
r_{L}(\mathcal{M}_{k,u}^{\alpha }) & \leq & \left[ 2^{3\cdot
k+1}-2^{k+r}\left( 2^{r}+2^{k}\right) \right] +\left[ 2^{2\cdot k}r_{H}(
\mathcal{M}_{r,u}^{\alpha ,2})+2^{k}r_{L}(\mathcal{M}_{r,u}^{\alpha ,4})
\right]
\end{array}
$
\item $
\begin{array}{ccc}
r_{E}(\mathcal{M}_{k,u}^{\alpha }) & \leq & \frac{11}{6}\left[ 2^{3\cdot
k+2}-2^{k+r}\left( 2^{r}+3\cdot 2^{k}\right) \right] +\left[ 2^{2\cdot
k}r_{H}(\mathcal{M}_{r,u}^{\alpha ,2})+2^{k}r_{E}(\mathcal{M}_{r,u}^{\alpha
,4})\right]%
\end{array}%
$
\end{enumerate}
\end{thm}

\pf
\[
\begin{array}{ccc}
r_{L}(\mathcal{M}_{k,u}^{\alpha }) & \leq & r_{L}(2^{2\cdot k}\mathcal{M}
_{k,u}^{\alpha ,2})+r_{L}(2^{k}\mathcal{M}_{k,u}^{\alpha ,4}) \\
& \leq & 2^{2\cdot k}r_{L}(\mathcal{M}_{k,u}^{\alpha ,2})+2^{k}r_{L}(
\mathcal{M}_{k,u}^{\alpha ,4}) \\
& \leq & 2^{2\cdot k}r_{H}(\mathcal{M}_{k,u}^{\alpha ,2})+2^{k}r_{L}(
\mathcal{M}_{k,u}^{\alpha ,4}) \\
& \leq & 2^{k}\left( 2^{2\cdot k}-2^{2\cdot r}\right) +2^{k}r_{L}(\mathcal{M}
_{r,u}^{\alpha ,4})+2^{2\cdot k}\left( 2^{k}-2^{r}\right) +2^{2\cdot k}r_{H}(
\mathcal{M}_{r,u}^{\alpha ,2}) \\
& \leq & \left[ 2^{3\cdot k+1}-2^{k+r}\left( 2^{r}+2^{k}\right) \right] +
\left[ 2^{2\cdot k}r_{H}(\mathcal{M}_{r,u}^{\alpha ,2})+2^{k}r_{L}(\mathcal{M
}_{r,u}^{\alpha ,4})\right]
\end{array}
\]
\qed

Similar arguments holds for $r_{E}(\mathcal{M}_{k,u}^{\alpha }).$
Similarly using the matrix \ref{equation:5}, Proposition \ref{prop:3} and Theorem \ref{thm:4} following
bounds can be obtained MacDonald code of type $\beta .$
\begin{thm}
for $
\begin{array}{ccccc}
u & \leq & r & \leq & k
\end{array}
$
\begin{enumerate}
\item $
\begin{array}{cccc}
r_{L}(\mathcal{M}_{k,u}^{\beta }) & \leq & 2^{2\cdot k-1}\left( 3\cdot
2^{k}-1\right) -2^{k+r-1}\left( 2^{k-1}+2^{r}-1\right) & +\left[ 2^{2\cdot
k}r_{H}(\mathcal{M}_{r,u}^{\beta ,2})+2^{k}r_{L}(\mathcal{M}_{r,u}^{\beta
,4})\right]
\end{array}
$
\item $
\begin{array}{ccc}
r_{E}(\mathcal{M}_{k,u}^{\alpha }) & \leq & \frac{11}{6}\left[ 2^{3\cdot
k+2}-2^{k+r}\left( 2^{r}+3\cdot 2^{k}\right) \right] +\left[ 2^{2\cdot
k}r_{H}(\mathcal{M}_{r,u}^{\alpha ,2})+2^{k}r_{E}(\mathcal{M}_{r,u}^{\alpha
,4})\right]
\end{array}
$
\end{enumerate}
\end{thm}
\pf
ByTheorem \ref{thm:4}
\[
\begin{array}{ccc}
r_{L}(\mathcal{M}_{k,u}^{\beta }) & \leq & r_{L}(2^{2\cdot k}\mathcal{M}
_{k,u}^{\beta ,2})+r_{L}(2^{k}\mathcal{M}_{k,u}^{\beta ,4}) \\
& \leq & 2^{2\cdot k}r_{L}(\mathcal{M}_{k,u}^{\beta ,2})+2^{k}r_{L}(\mathcal{
M}_{k,u}^{\beta ,4}) \\
& \leq & 2^{2\cdot k}r_{H}(\mathcal{M}_{k,u}^{\beta ,2})+2^{k}r_{L}(\mathcal{
M}_{k,u}^{\beta ,4}) \\
& \leq &
\begin{array}{c}
2^{k}\left[ 2^{k-1}\left( 2^{k}-1\right) -2^{r-1}\left( 2^{r}-1\right)
\right] +2^{k}r_{L}(\mathcal{M}_{r,u}^{\beta ,4}) \\
+2^{k}\left( 2^{k}-2^{u}\right) +2^{2\cdot k}r_{H}(\mathcal{M}_{r,u}^{\beta
,2})
\end{array}
\\
& \leq &
\begin{array}{c}
\left[ 2^{2\cdot k-1}\left( 3\cdot 2^{k}-1\right) -2^{k+r-1}\left(
2^{k-1}+2^{r}-1\right) \right] \\
+\left[ 2^{2\cdot k}r_{H}(\mathcal{M}_{r,u}^{\beta ,2})+2^{k}r_{L}(\mathcal{M
}_{r,u}^{\beta ,4})\right]
\end{array}
\end{array}
\]
\qed
Similar arguments holds for $r_{E}(\mathcal{M}_{k,u}^{\beta }).$
\begin{thm}
 $r_{L}(\mathcal{M}_{k,u}^{\alpha^{\bot} })=r_{L}(\mathcal{M}_{k,u}^{\beta^{\bot} })=2$,
  $r_{E}(\mathcal{M}_{k,u}^{\alpha^{\bot} })\leq 6$ and $r_{E}(\mathcal{M}_{k,u}^{\beta^{\bot} })\leq 6 $
\end{thm}
\pf
The same proof such as the theorem \ref{thm:14}.
\qed
\section{$\mathbb{Z}_{2}\mathbb{Z}_{4}-$Reed-Muller Code}
\bigskip In \cite{Ronquillo} and \cite{Rif} the additive Reed-Muller codes over $\mathbb{Z}_{2}\mathbb{Z}_{4}$ is known by $\mathcal{A}
\mathcal{R}\mathcal{M}\left(r,m \right)$. Just as there is only one $\mathcal{R}
\mathcal{M}\left(r,m \right) $ family in the binary case, in the $\mathbb{Z}_{2}\mathbb{Z}_{4}-$
additive case there are $\lfloor\dfrac{m+2}{2}\rfloor$ families for each value of $m$. Each one
of these families will contain any of the $\lfloor\dfrac{m+2}{2}\rfloor$ non-isomorphic
$\mathbb{Z}_{2}\mathbb{Z}_{4}-$linear extended perfect codes which are known to
exist for any $m$ (see \cite{Rifa}). We will identify each family $\mathcal{A}
\mathcal{R}\mathcal{M}_{s}\left(r,m \right)$ by a subindex $s\in\left\lbrace0,\cdots,
\lfloor\dfrac{m}{2}\rfloor \right\rbrace $. As in linear case we have the  generator matrix
for the code $\mathcal{A}\mathcal{R}\mathcal{M}\left(r,m \right)$ is given by
\begin{center}
$G\left(r,m \right) =\left[
\begin{tabular}{l|l}
$G\left(r,m-1 \right)$ & $G\left(r,m-1 \right)$ \\ \hline
$0$ & $G\left(r-1,m-1 \right)$
\end{tabular}
\right] $
\end{center}
\begin{prop} \cite{Rif}
The additive Reed-Muller Code $\mathcal{A}\mathcal{R}\mathcal{M}\left(r,m \right)$ of order $r$
has the following properties
\item[1] Minimum distance $d=2^{m-r}$
\item[2] If $k=\sum_{i=0}^{r}\left(
\begin{array}{c} m\\ i
\end{array} \right) $ then $\vert\mathcal{A}\mathcal{R}\mathcal{M}\left(r,m \right)\vert=2^{k}$
\item[3] There exists a coordinate permutation $\sigma_{r}$ such that
$\sigma_{r}\left( \mathcal{A}\mathcal{R}\mathcal{M}\left(r-1,m \right)\right)\subset
\mathcal{A}\mathcal{R}\mathcal{M}\left(r,m \right)$, for $r>0$
\end{prop}

\bigskip In this section we give the covering radius of first order Reed Muller
code over $\mathbb{Z}_{2}\mathbb{Z}_{4}$. Let $2\leq i\leq m-1$. Let $v_{i}$ be a vector of length $
2^{m-1}$ consisting of successive blocks of $00$'s and $11$'s each of size $
2^{\left( m-1\right) -i}$ and if $\overline{11}=\left( 11\text{ }11\cdots
11\right) \in \left( \mathbb{Z}_{2}\mathbb{Z}_{4}\right) ^{2^{m-1}}.$ Let $G$
be a $2^{m-1}\times \left( m-1 \right) $ matrix given by
\begin{equation*}
G\left( 1,m-1 \right)=\left(
\begin{array}{cccccccccc}
00 & 00 & \cdots & 00 & 00 & 02 & 02 & \cdots & 02 & 02 \\
\vdots & \vdots & \ddots & \vdots & \vdots & \vdots & \vdots & \ddots &
\vdots & \vdots \\
00 & 02 & \cdots & 00 & 02 & 00 & 02 & \cdots & 00 & 02 \\
11 & 11 & \cdots & 11 & 11 & 11 & 11 & \cdots & 11 & 11
\end{array}%
\right)
\end{equation*}
The matrix $ G\left( 1,m-1 \right) $ is also takes the form
\begin{equation*}
\left(
\begin{tabular}{llllll}
&  &  & \multicolumn{1}{|l}{} &  &  \\
& $0_{2^{m-2}\times \left( m-2\right) }$ &  & \multicolumn{1}{|l}{} & $
2\cdot S_{2,m-2}^{\alpha }$ &  \\
&  &  & \multicolumn{1}{|l}{} &  &  \\ \hline
$11$ & $\cdots $ & $11$ & $11$ & $\cdots $ & $11$
\end{tabular}
\right),
\end{equation*}
where $S_{2,m-2}^{\alpha }$ is a binary simplex code of type $\alpha$.

The code generated by $G\left( 1,m-1 \right)$ is called the first order Reed-Muller code over $
\mathbb{Z}_{2}\mathbb{Z}_{4}$, denoted $\mathcal{A}\mathcal{R}\mathcal{M}\left(1,m-1 \right)$. It is a linear
code over $\mathbb{Z}_{2}\mathbb{Z}_{4}$ \cite{Manish}.
\begin{thm}
If C is the code generated by G then $r_{L}\left( C\right) = r_{E}\left( C\right) =2^{m-1}.$
\end{thm}

\pf
Let $x=1111\cdots11 \in \left( \mathbb{Z}_{2}\mathbb{Z}_{4}\right)^{2^{m-1}} $, then
$d_{L}\left(x,C\right)=d_{E}\left(x,C\right) =2^{m-1}  $.
\qed

\section{The Binary Gray Images of Simplex and MacDonald Codes over $\mathbb{Z}_{2}\mathbb{Z}_{4} $}
\bigskip The binary version of $S _{k}^{\alpha }$ a simplex Codes over $\mathbb{Z}_{2}\mathbb{Z}_{4} $ is
given by the following theorem.

\begin{thm}\label{thm:7}
Let $S _{k}^{\alpha }$ is a $\mathbb{Z}_{2}\mathbb{Z}_{4}-$simplex code of type $\alpha$ of minimum Lee weight $d_{L}$, then $\Phi  _{L}(S _{k}^{\alpha})$ is
a concatenation of $2^{2k}\left( 2^{k}+1\right) $ binary simplex code with parameters\\
 $[2^{3k}\left( 2^{k}+1\right);k;d_{H}]$.
\end{thm}
\pf
If $\Theta_{k}^{\alpha }$ is generator matrix of $\mathbb{Z}_{2}\mathbb{Z}_{4}-$simplex Codes $
S_{k}^{\alpha }$. Then $\Phi _{L}(\Theta_{k}^{\alpha })$ is in the form
\begin{equation*}
\Phi _{L}(\Theta_{k}^{\alpha })=\left( \overset{2^{2k}\left( 2^{k}+1\right)}{\overbrace{
\begin{tabular}{l|l|l|l}
$m_{k}$ & $m_{k}$ & $\cdots $ & $m_{k}$
\end{tabular}
}}\right)
\end{equation*}
where $m_{k}$ is generator matrix of binary simplex code $S_{k}$. The result
follows obtained by induction on $k$.
\qed
The binary version of $S _{k}^{\beta }$ a simplex Codes over $\mathbb{Z}_{2}\mathbb{Z}_{4} $ is
given by the following theorem.

\begin{thm}\label{thm:8}
Let $S _{k}^{\beta}$ is a $\mathbb{Z}_{2}\mathbb{Z}_{4}-$simplex code of
type $\beta$ of minimum Lee weight $d_{L}$, then $\Phi  _{L}(S _{k}^{\beta})$
is a concatenation of $2^{k}\left( 2^{k-1}+1\right)  $ binary simplex code with parameters\\
$[2^{k}\left( 2^{k-1}+1\right)\left(2^{k} -1\right);k;d_{H}]$.
\end{thm}
\pf
Same as the proof in theorem \ref{thm:7}
\qed
By analogy the binary images of $\mathcal{M}_{k,u}^{\alpha }$ (resp., $\mathcal{M}_{k,u}^{\beta }$) a MacDonald and Reed-Muller Codes over
$\mathbb{Z}_{2}\mathbb{Z}_{4} $ is given by the following theorem.

\begin{thm}
Let $\mathcal{M}_{k,u}^{\alpha }$ (resp.,$\mathcal{M}_{k,u}^{\beta }$) are
a $\mathbb{Z}_{2}\mathbb{Z}_{4}-$simplex code of type $\alpha$ and $\beta$ of
minimum Lee weight $d_{L}$, then $\Phi  _{L}(\mathcal{M}_{k,u}^{\alpha })$
(resp.,$\Phi  _{L}(\mathcal{M}_{k,u}^{\beta })$)
is a concatenation of $2^{2\left( k-1\right) }\left( 2^{k-1}+1\right)$ (resp.,
 $2^{2(k-1)}\left(2^{k}-1 \right) $) binary simplex code with parameters
$[2^{2\left( k-1\right) }\left( 2^{k-1}+1\right)\left(2^{k}-2^{u} \right)  ;k;d_{H}]$
(resp.,$[2^{2(k-1)}\left(2^{k}-1 \right)\left(2^{k}-2^{u} \right)  ;k;d_{H}]$).
\end{thm}

\pf
by Theorems \ref{thm:7} and \ref{thm:8} the results are easy.
\qed

\begin{thm}
The additive first order Reed-Muller code $\mathcal{A}\mathcal{R}\mathcal{M}\left(1,m-1 \right)$
 over $\mathbb{Z}_{2}\mathbb{Z}_{4} $of minimum Lee weight $d_{L}$, to a binary image
 $\Phi  _{L}(\mathcal{A}\mathcal{R}\mathcal{M}\left(1,m-1 \right))$ is a code with parameters
 $\left[3\cdot2^{m-1};m-1;d_{H}=2^{m-2} \right] $
\end{thm}

\pf
If $G\left(1,m-1 \right)$ is the matrix of the additive first order Reed-Muller code $\mathcal{A}\mathcal{R}\mathcal{M}\left(1,m-1 \right)$
 over $\mathbb{Z}_{2}\mathbb{Z}_{4} $, then the image under the Gray map is a binary code of  generator matrix given by
 \begin{center}
 $\Phi  _{L}(G\left(1,m-1 \right))=\left(
\begin{tabular}{lll|lll|lll}
&  &  &  &  &  &  &  &  \\
& $0_{2^{m-2}\times \left( m-2\right) }$ &  &  & $\overline{\mathcal{G}\left(
1,m-2\right) }$ &  &  & $\mathcal{G}\left( 1,m-2\right) $ &  \\
&  &  &  &  &  &  &  &  \\ \hline
$11$ & \multicolumn{1}{l}{$\cdots $} & \multicolumn{1}{l|}{$11$} & $00$ &
{$\cdots $} & \multicolumn{1}{l|}{$00$} & $11$ & $\cdots
$ & $11$%
\end{tabular}%
\right) $
 \end{center}, where $\mathcal{G}\left( 1,m-2\right) $ is a matrix of binary first order Reed-Muller code.
 For the minimum distance if takes two vectors $u$ and $v$ in $\mathcal{A}\mathcal{R}\mathcal{M}\left(1,m-1 \right)$, we have $d_{H}=2^{m-2} $.

\qed

\paragraph{Conclusion}
In this paper, we have computed bounds on the covering radius of Simplex and
MacDonald codes of type $\alpha $ and $\beta $ over $\mathbb{Z}_{2}\mathbb{Z}
_{4}$ and also provided exact values in some cases.Of course, another
direction and interesting research in this topic is the computed bounds on
the covering radius of Simplex and MacDonald codes of type $\alpha $ and $
\beta $ over other ring.

\end{document}